\documentclass[aps,prl,twocolumn,showpacs]{revtex4-1}

\usepackage{amsmath}
\usepackage{amssymb}
\usepackage{graphicx}
\usepackage{epstopdf}

\newcommand{\sech}{\mathrm{sech}}

\begin{document}
\title{Mechanical Actuation of Magnetic Domain-Wall Motion}

\author{Se Kwon Kim}
\author{Daniel Hill}
\author{Yaroslav Tserkovnyak}
\affiliation{Department of Physics and Astronomy, University of California, Los Angeles, California 90095, USA}

\date{\today}

\begin{abstract}
We theoretically study the motion of a magnetic domain wall induced by transverse elastic waves in a one-dimensional magnetic wire, which respects both rotational and translational symmetries. By invoking the conservation of the associated total angular and linear momenta, we are able to derive the torque and the force on the domain wall exerted by the waves. We then show how ferromagnetic and antiferromagnetic domain walls can be driven by circularly- and linear-polarized waves, respectively. We envision that elastic waves may provide effective means to drive the dynamics of magnetic solitons in insulators.
\end{abstract}

\pacs{75.78.-n, 75.60.Ch, 62.30.+d, 63.20.-e}

\maketitle

\emph{Introduction.}|Phonons, quanta of elastic vibrations, are ubiquitous in condensed matter systems including magnets. Owing to their gapless nature, they can easily absorb energy from excited spins, thereby engendering the damping term in the description of spin dynamics \cite{GilbertIEEE2004}. Apart from this passive role, the idea of actively using phonons to induce magnetic dynamics has been recently gaining attention in spintronics. It has been experimentally demonstrated that acoustic pulses can induce coherent magnetization precession \cite{ScherbakovPRL2010, *KimPRL2012} via spin-lattice coupling \cite{KittelPR1958}. Also excitation of elastic waves can generate spin currents \cite{UchidaNM2011, *WeilerPRL2012} and thereby drive magnetic bubbles \cite{OgawaPNAS2015}.

A domain wall in an easy-axis magnet is one of the simplest and well-studied topological solitons \cite{*[][{, and references therein.}] KosevichPR1990}, which has practical importance exemplified by the racetrack memory \cite{ParkinScience2008}. They can be driven by various means: a magnetic field \cite{SchryerJAP1974}, an electric field \cite{*[][{, and references therein.}] PyatakovPU2015}, a spin-polarized electric current \cite{SlonczewskiJMMM1996, *BergerPRB1996, *SwavingPRB2011, *HalsPRL2011}, a temperature gradient \cite{KovalevEPL2012, *JiangPRL2013}, or a spin wave \cite{YanPRL2011, *TvetenPRL2014}. Moving domain walls have been known to generate and drag phonons, which in turn gives rise to the damping force on the walls \cite{*[][{, and references therein.}] BaryakhtarSPU1985}. This force increases as the domain wall approaches the speed of sound, which was pointed out as the origin of the plateau in the dependence of the domain-wall speed on an external field \cite{DemokritovJMMM1991}.

\begin{figure}
\includegraphics[width = \columnwidth]{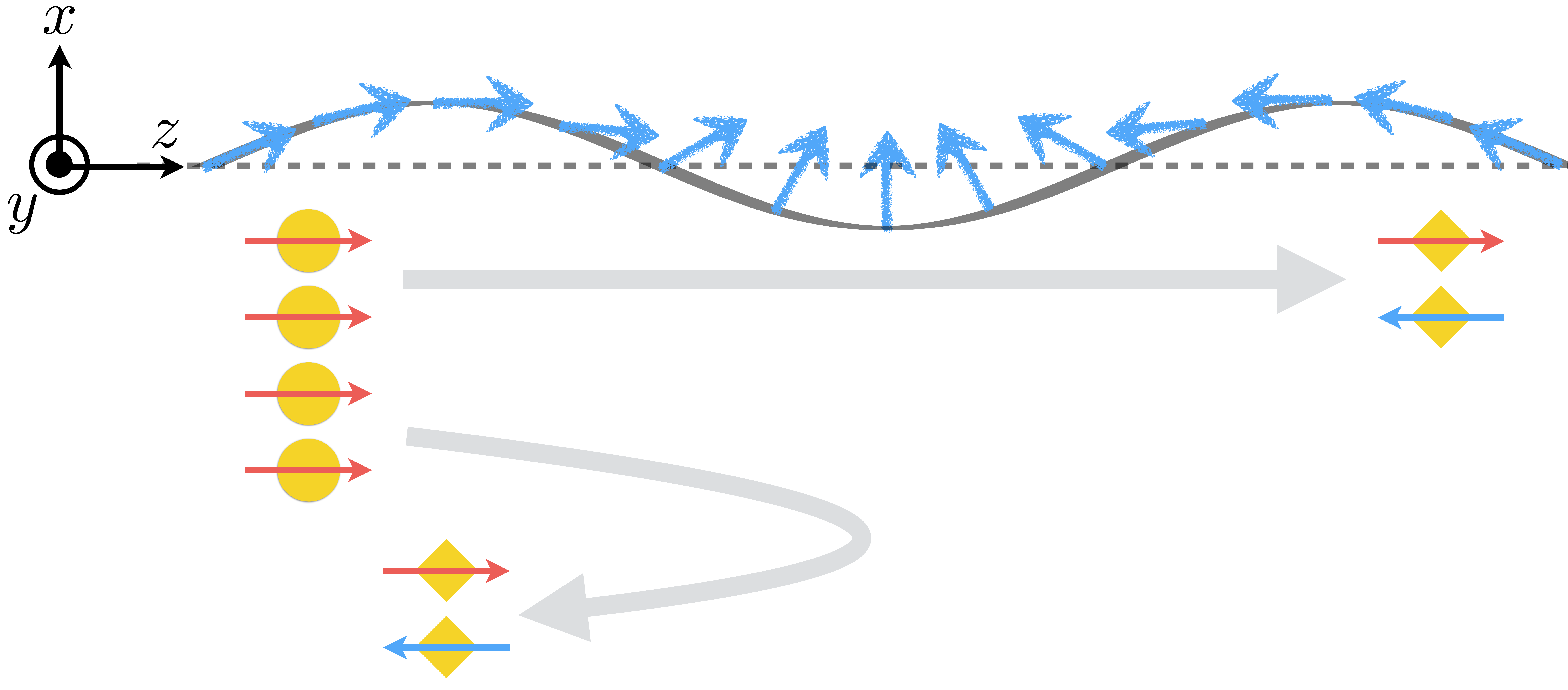}
\caption{A schematic illustration of a ferromagnetic wire with a magnetic domain wall (shown by blue thick arrows) and a transverse elastic deformation (which is exaggerated for illustrative purposes). The yellow circles represent incoming phonons, quanta of elastic waves; the yellow diamonds represents transmitted and reflected phonons. The red (blue) arrows on circles represent phonons' angular momentum in the positive (negative) $z$ direction. Phonons are injected from the left; some of them are reflected by the domain wall and thereby exert the force on it;  some of the transmitted and reflected phonons change their angular momentum and thereby exert the torque on the wall.}
\label{fig:fig1}
\end{figure}

In this Letter, we study the reciprocal problem: actuation of the magnetic domain-wall motion via the phonon current, which can be injected by mechanical means. The stress-induced motion of a domain wall has been previously studied in Ref.~\cite{BryanPRB2012}, in which the domain wall is energetically driven by the axial stress gradient generated by the static voltage profile in piezoelectric materials. Differing from that, we focus on the effects of the dynamic phonon current on the domain wall via scattering. Specifically, we consider a one-dimensional magnetic wire with a coaxial easy-axis anisotropy, which can be realized by a single-crystalline iron nanowire embedded in a carbon nanotube \cite{LipertAPL2010}. It respects the rotational and translational symmetries and thus conserves the total angular and linear momenta. A magnetic domain wall breaks both symmetries, which opens channels for the exchange of both momenta with phonons. See Fig.~\ref{fig:fig1} for an illustration of a domain-wall configuration for a ferromagnetic system. We show that the domain wall is birefringent for transverse waves and can thus act as a waveplate that alters the circular polarization---and thus the angular momentum---of phonons traveling through it. This change of phonons' angular momentum applies the torque on the domain wall. Reflection of phonons by the domain wall gives rise to the force acting on it. We study the domain-wall motion induced by the phononic torque and the force in ferromagnets and antiferromagnets.

\emph{Main Results.}|Our model system is a one-dimensional magnetic wire stretched along the global $z$ axis by an external tension, in which the magnetic order parameter tends to align with the local orientation of the wire. The order parameter is the local spin angular-momentum density for ferromagnets and the local N{\'e}el order for antiferromagnets. For temperatures well below the ordering temperature, the local order parameter has the saturated magnitude and thus can be represented by the unit vector $\mathbf{n} (\zeta, t)$ pointing along its direction. Here, $\zeta$ is the internal coordinate of the lattice atoms along the wire. In this Letter, we are interested in the interaction between the magnetic soliton---domain wall---and the transverse vibrations of the wire, which are represented by $u(\zeta, t)$ and $v(\zeta, t)$ for the displacements of the atom at $\zeta$ in the lab-frame $x$ and $y$ direction, respectively \footnote{By assuming that the Young's modulus is much larger than the applied tension, $E \gg \mathcal{T}$, we shall neglect longitudinal displacements by focusing on low-energy transverse modes \cite{Kardar, Goldstein, ButikovPS2012}.}. We shall focus on small displacements by working to the quadratic order in $u$ and $v$. In studying the dynamics of elastic waves, we shall assume that the dynamics of the magnetization is slow enough to be treated as static in the equations of motion for elastic waves, which would be valid if the speed of sound is much larger than that of magnons.

The potential energy that involves the magnetic order parameter is given by
\begin{equation}
U_\text{m} = \int d\zeta \, \left[ A \mathbf{n}'^2 + K \{ 1 - (\mathbf{n} \cdot \mathbf{t})^2 \} \right] / 2 \, , \\
\label{eq:Um}
\end{equation}
where the positive constants $A$ and $K$ are the exchange and anisotropy coefficients, respectively \footnote{Transverse elastic deformations may affect the magnetic potential energy by modifying the geometry of the wire, which can be captured by an additional potential-energy term $\delta U_\text{m} = \int d\zeta \, (u'^2 + v'^2) [ \xi A \mathbf{n}'^2 + \nu K \{ 1 - (\mathbf{n} \cdot \mathbf{t})^2 \} ] / 2$, where $\xi$ and $\nu$ are the dimensionless parameters. This term modifies the strength of the potential for elastic waves in Eq.~(\ref{eq:de}), $2 \tilde{\kappa} \mapsto [2 - (\xi + \nu)] \tilde{\kappa}$ for $u$ and $\tilde{\kappa} \mapsto [1 - (\xi + \nu)] \tilde{\kappa}$ for $v$, which does not change the elastic-wave-induced motion of the domain wall qualitatively.}. Here, $'$ is the derivative with respect to the intrinsic coordinate $\zeta$; $\mathbf{t} (\zeta, t) \equiv (u', v', \sqrt{1 - u'^2 - v'^2})$ is the unit tangent vector of the wire. The magnetic anisotropy can be rooted in either the magneto-crystalline anisotropy or the shape anisotropy induced by dipolar interactions. When the wire is straight along the $z$ axis, $u \equiv v \equiv 0$, there are two ground states: $\mathbf{n} \equiv \hat{\mathbf{z}}$ and $\mathbf{n} \equiv - \hat{\mathbf{z}}$. A domain wall is a stationary solution of $\delta_\mathbf{n} U_\text{m} = 0$ that interpolates two ground states $\mathbf{n} (\zeta = \pm \infty) = \mp \hat{\mathbf{z}}$. It is given by \cite{SchryerJAP1974}
\begin{subequations}
\label{eq:dw}
\begin{align}
n_x (\zeta) &= \sech [ (\zeta - Z) / \lambda ] \cos \Phi \, , \\
n_y (\zeta) &= \sech [ (\zeta - Z) / \lambda ] \sin \Phi \, , \\
n_z (\zeta) &= - \tanh \left[ (\zeta - Z) / \lambda \right] \, .
\end{align}
\end{subequations}
Here, $Z$ and $\Phi$ are the position and the azimuthal angle of the domain wall, respectively; $\lambda \equiv \sqrt{A / K}$ is the characteristic length scale of the problem, corresponding to the domain-wall width. $Z$ and $\Phi$ parametrize two zero modes of the domain wall, which are associated with the breaking of the translational and spin-rotational symmetries. The dynamics of the position $Z$ induced by the wire's transverse vibrations is of our main interest.

The linearized dynamics of the transverse displacements of the stretched wire can be described by the Lagrangian \cite{Kardar}
\begin{equation}
L_\text{e} = \int d\zeta \, \left[ \mu (\dot{u}^2 + \dot{v}^2) - \mathcal{T} (u'^2 + v'^2) \right] / 2 \, ,
\label{eq:Le}
\end{equation}
where the positive constants $\mu$ and $\mathcal{T}$ are the mass density of the wire and the applied tension, respectively \footnote{We neglect the bending energy $\propto (u'')^2$ \cite{CL} by assuming that the tension is strong so that the potential energy $\propto \mathcal{T}$ in $L_\text{e}$ dominates over the bending energy.}. The equations of motion for $u$ and $v$, that are derived from the Lagrangian $L_\text{e}$ in conjunction with the potential energy $U_\text{m}$ [Eq.~(\ref{eq:Um})], are given by
\begin{subequations}
\label{eq:uv}
\begin{align}
\mu \ddot{u} - \left[ \left\{ \mathcal{T} + K (n_z^2 - n_x^2) \right\} u' \right]' &= - K (n_z n _x)' \, ,  \\ 
\mu \ddot{v} - \left[ \left\{ \mathcal{T} + K (n_z^2 - n_y^2) \right\} v' \right]' &= - K (n_z n _y)' \, . 
\end{align}
\end{subequations}
For the uniform ground states, $\mathbf{n} \equiv \pm \hat{\mathbf{z}}$, the right-hand sides vanish and the tension is effectively increased from $\mathcal{T}$ to $\mathcal{T}_\kappa \equiv (1 + \kappa) \mathcal{T}$ with $\kappa \equiv K / \mathcal{T}$. The dispersion relation is given by $\omega = \pm v_0 k$ with the speed $v_0 \equiv \sqrt{\mathcal{T}_\kappa / \mu}$. Using the propagating-wave solutions to the above equations in the presence of the domain wall, details of which will be shown later, we can derive the phononic torque and the force on the wall by invoking the conservation of the angular and linear momenta. The induced domain-wall speed is quadratic in the amplitude of waves, which allows us to assume that the domain wall is static in Eqs.~(\ref{eq:uv}) to the linear order in the amplitude \footnote{In the quantum regime, the domain wall can be considered static only when its effective inertia is much larger than that of incoming phonons.}.

\begin{figure}
\includegraphics[width = 0.9 \columnwidth]{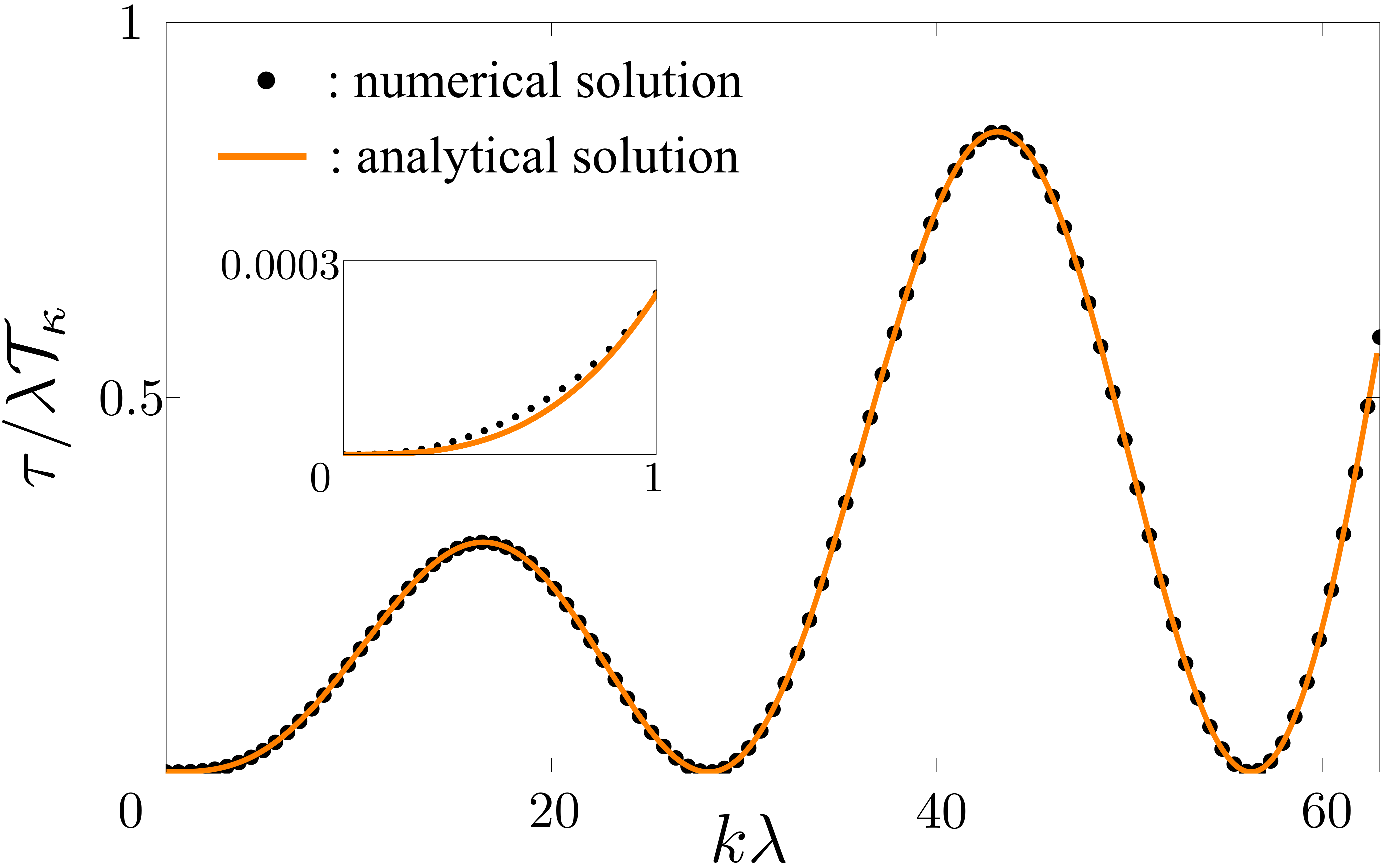}
\caption{The torque $\tau$ on the domain wall by the circularly-polarized waves as a function of the wavenumber $k \lambda$ for the parameters $\kappa = 0.2$ and $a = \lambda / 10$. The solid line is obtained with the analytical expression for $\tau$ in Eq.~(\ref{eq:tau-cp}); the dots are obtained with $\tau$ in Eq.~(\ref{eq:tau}) calculated from numerical solutions of the differential equations~(\ref{eq:de}). The inset shows a zoom-in at small wave vectors $k \lambda < 1$.}
\label{fig:fig2}
\end{figure}



First, circularly-polarized waves incoming from the left, $u(\zeta, t) = a \cos (k \zeta - \omega t)$ and $v(\zeta, t) = - a \sin(k \zeta - \omega t)$, exert the torque (i.e., the transfer of angular momentum) on the domain wall, 
\begin{equation}
\tau \simeq \mathcal{T}_\kappa a^2 k [1 - \cos \{ k \lambda \ln (1 - \kappa) \} ] \, , 
\label{eq:tau-cp}
\end{equation}
for high-energy waves $k \lambda \gg 1$, which is obtained by the subtraction of the angular momentum current of the transmitted wave, $\mathcal{T}_\kappa a^2 k \cos \{ k \lambda \ln (1 - \kappa) \}$, from that of the incoming wave, $\mathcal{T}_\kappa a^2 k$. The physical origin of the torque can be understood as follows. From Eqs.~(\ref{eq:uv}), the domain wall locally modifies the tension for the $u$ and $v$ displacements by $K [1 - 2 \sech^2 (\zeta / \lambda)]$ and $K [1 - \sech^2 (\zeta / \lambda)]$, respectively. The $v$ component thus propagates faster than the $u$ component within the domain wall, which acts as a birefringent medium that can alter the polarization of the wave. The argument of the cosine function in Eq.~(\ref{eq:tau-cp}) is the relative phase shift of $u$ and $v$ components of the transmitted wave, $\phi_{u, t} - \phi_{v, t} \simeq k \lambda \ln (1 - \kappa)$. Figure~\ref{fig:fig2} shows the torque $\tau$ as a function of the wavenumber $k \lambda$. Note that it oscillates as a function of $k \lambda$ with the period of $2 \pi / \ln(1 - \kappa)$. This torque by the elastic waves can drive ferromagnetic domain walls, analogous to the torque of spin waves \cite{YanPRL2011}. The steady-state speed of the ferromagnetic domain wall is $V = \tau / 2 s$ [Eq.~(\ref{eq:V-fm})] in the absence of damping, where $s \equiv \hbar S / \mathcal{V}$ is the saturated spin density ($\mathcal{V}$ is the volume per spin).

\begin{figure}
\includegraphics[width = 0.9 \columnwidth]{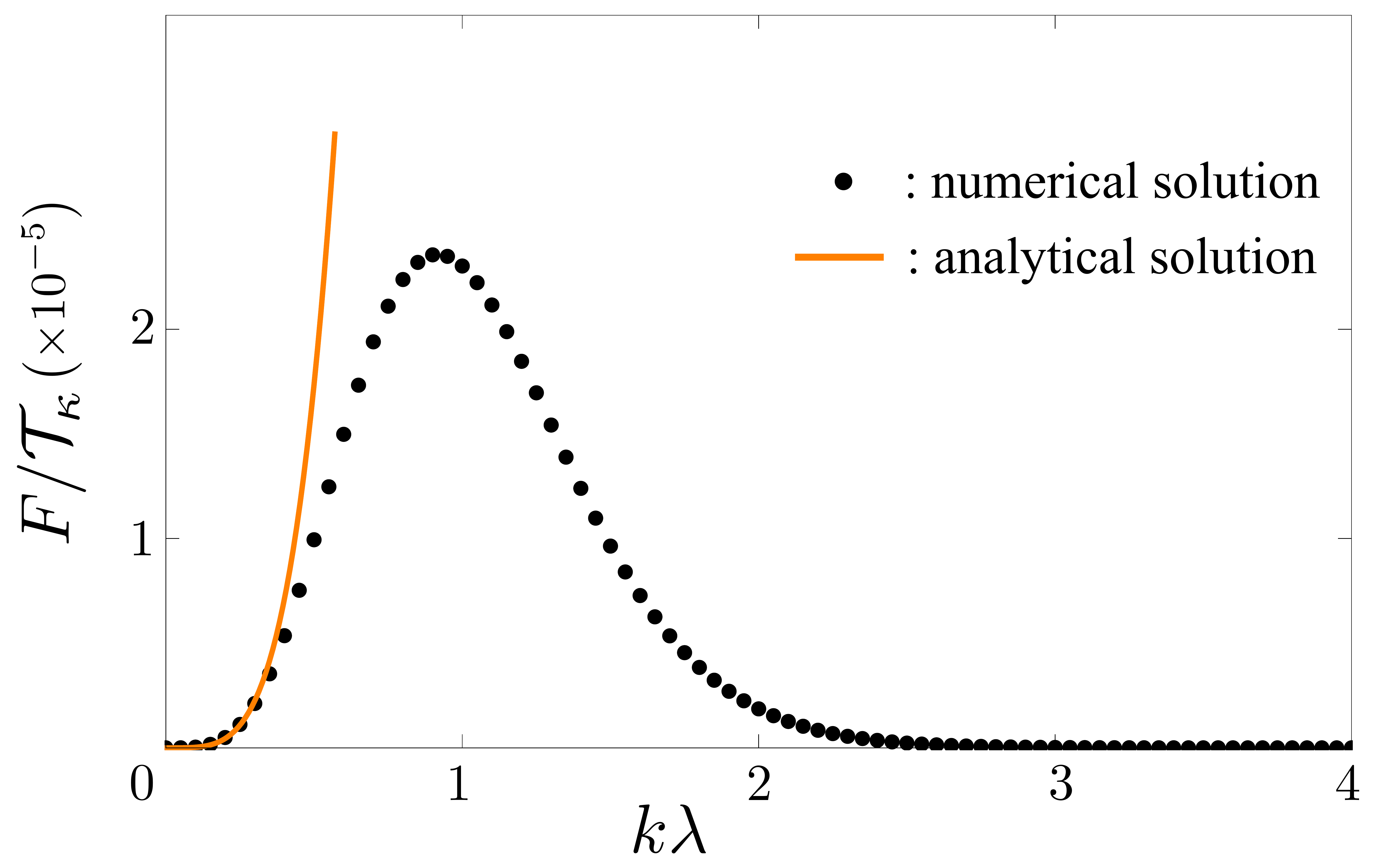}
\caption{The force $F$ on the domain wall exerted by the linearly-polarized waves perpendicular to the wall plane for the parameters $\kappa = 0.2$ and $a = \lambda / 10$. The solid line is obtained with the analytical expression for $F$ in Eq.~(\ref{eq:F-lp}); the dots are obtained with $F$ in Eq.~(\ref{eq:F}) calculated from numerical solutions of the differential equations~(\ref{eq:de}).}
\label{fig:fig3}
\end{figure}

Secondly, linearly-polarized waves incoming from the left, $v(\zeta, t) = a \cos (k \zeta - \omega t)$ and $u(\zeta, t) \equiv 0$, exert no torque, but a finite force (i.e., the transfer of linear momentum) on the domain wall due to the reflection, 
\begin{equation}
F \simeq \mathcal{T}_\kappa \kappa^2 \lambda^2 a^2 k^4 \, ,
\label{eq:F-lp}
\end{equation}
for low-energy waves $k \lambda \ll 1$. It is the product of the pressure (i.e., the linear momentum current) of the incoming wave, $\mathcal{T}_\kappa a^2 k^2 / 2$, and twice the reflection probability, $2 \kappa^2 \lambda^2 k^2$. The reflection probability is exponentially small for high-energy waves $k \lambda \gg 1$, and so is the force. Figure~\ref{fig:fig3} shows the force $F$ as a function of the wavenumber $k \lambda$. This force of the elastic waves can drive antiferromagnetic domain walls, analogous to the force of spin waves \cite{TvetenPRL2014, KimPRB2014}. The steady-state speed of the antiferromagnetic domain wall is $V = \lambda F / 2 \alpha s$ [Eq.~(\ref{eq:V-afm})], where $\alpha$ is the Gilbert damping constant.

\emph{Transverse waves.}|Let us solve the differential equations~(\ref{eq:uv}) for $u$ and $v$ in the presence of the static domain wall given by Eqs.~(\ref{eq:dw}). A general solution is composed of static and dynamic components. A static one is determined by the right-hand sides of the equations \cite{BaryakhtarSPU1985, *ZvezdinJETP1992}, whereas dynamic components, which are of interest to us, are the propagating waves, for which we can neglect the right-hand sides. For the monochromatic solutions, i.e., $\propto \exp(- i \omega t)$, the equations are given by
\begin{subequations}
\label{eq:de}
\begin{align}
\left[ \left\{ 1 - 2 \tilde{\kappa} \, \sech^2 (\zeta / \lambda) ) \right\} u' \right]' = - k^2 u \, , \\
\left[ \left\{ 1 - \tilde{\kappa} \, \sech^2 (\zeta / \lambda) \right\} v' \right]' = - k^2 v  \, ,
\end{align}
\end{subequations}
with $k^2 \equiv \omega^2 / v_0^2$ and $\tilde{\kappa} \equiv \kappa / (1 + \kappa)$. We shall focus on solutions for $v$ henceforth, from which we can obtain solutions for $u$ by replacing $\tilde{\kappa}$ by $2 \tilde{\kappa}$. For the given incoming-wave component, the solution far away from the wall can be characterized by four real numbers: the amplitude $t > 0$ and the phase-shift $\phi_t$ of the transmitted component and the amplitude $r > 0$ and the phase-shift $\phi_r$ of the reflected component:
\begin{equation}
v (\zeta) \propto \begin{cases}
e^{i k \zeta} + r e^{- i k \zeta + i \phi_r} \, , & \text{ for } \zeta \ll - \lambda \, \\
t e^{i k \zeta + i \phi_t} \, , & \text{ for } \zeta \gg \lambda \, .
\end{cases} 
\end{equation}
The equation can be transformed into a quantum-mechanic scattering problem by introducing a new coordinate $\eta$ satisfying $d \zeta / d \eta = 1 - \tilde{\kappa} \, \sech^2 (\zeta / \lambda)$:
\begin{equation}
\left[- \frac{d^2}{d \eta^2}  + k^2 \tilde{\kappa} \, \sech^2 \frac{\zeta(\eta)}{\lambda} \right] v = k^2 v \, .
\end{equation}

We obtain approximate solutions in the two extreme energy regimes. First, in the high-energy limit, $k \lambda \gg 1$, we use the Wentzel-Kramers-Brillouin approximation \cite{LL3}, within which the solution is $v(\zeta) \propto \exp \left[ i k \int d\zeta \{ 1 - \tilde{\kappa} \, \sech^2 (\zeta / \lambda) \}^{-1/2} \right]$ in the original coordinate $\zeta$ with the transmission amplitude $t = 1$. The phase shift of the transmitted wave is given by
\begin{equation}
\phi_t = k \int_{-\infty}^{\infty} \frac{d \zeta}{\sqrt{1 - \tilde{\kappa} \, \sech^2 (\zeta / \lambda)}} = - k \lambda \ln (1 - \tilde{\kappa}) \, .
\end{equation}
In the low-energy limit, we approximate the potential by the delta-function barrier with the height $h_k \equiv 2 k^2 \sqrt{\tilde{\kappa} / (1 - \tilde{\kappa})} \arcsin \sqrt{\tilde{\kappa}}$ that is the spatial integral of the potential. After solving the scattering problem and going back to the original coordinate $\zeta$, we obtain
\begin{equation}
r = \tilde{\kappa} k \lambda \, , \quad \phi_r = - \pi / 2, \quad t = 1 \, , \quad \phi_t = \tilde{\kappa} k \lambda \, , 
\end{equation}
to the first order in $k \lambda$. Note that for a small anisotropy, $\tilde{\kappa} \ll 1$, the phase shifts of the transmitted wave in the two regimes coincide: $- k \lambda \ln(1 - \tilde{\kappa}) \simeq \tilde{\kappa} k \lambda$.

\emph{Torque and force.}|In the uniform state, $\mathbf{n} (\zeta) \equiv \hat{\mathbf{z}}$, the effective Lagrangian density for the waves which includes the effect of the anisotropy is given by 
$
\mathcal{L} = \mu (\dot{u}^2 + \dot{v}^2) - \mathcal{T}_\kappa (u'^2 + v'^2) \, .
$
Axial symmetry of the Lagrangian implies conservation of the corresponding angular momentum. The temporal and spatial components of the associated N{\"o}ther current \cite{Goldstein, KimPRB2014} are given by $\rho^s = \mu (u \dot{v} - v \dot{u})$ and $I^s = - \mathcal{T}_\kappa (u v' - v u')$, which are, respectively, the density and the current of the (orbital) angular momentum \footnote{\textcite{ZhangPRL2014} showed that phonons in magnetic crystals can have nonzero orbital angular momentum in equilibrium.}. We obtain the linear momentum density $T^{10} = - \mu \dot{u} u' + (u \rightarrow v)$ and the current $T^{11} = (\mu \dot{u}^2 + \mathcal{T}_\kappa u'^2) / 2 + (u \rightarrow v)$ from the stress-energy tensor, $T^{\alpha \beta} \equiv \partial^\alpha u [\partial \mathcal{L} / \partial (\partial_\beta u)] + (u \rightarrow v) - \delta^{\alpha \beta} \mathcal{L}$ \cite{Goldstein}. For monochromatic waves, $u(\zeta, t) = u_0 \cos(k \zeta - \omega t)$ and $v(\zeta, t) = v_0 \cos(k \zeta - \omega t + \Delta \phi)$, the angular and linear momentum currents are given by
\begin{equation}
I^s = \mathcal{T}_\kappa u_0 v_0 \sin (\Delta \phi) k \, , \quad T^{11} = \mathcal{T}_\kappa (u_0^2 + v_0^2)  k^2 / 2 \, .
\end{equation}



Let us now derive the torque and the force on the domain wall exerted by elastic waves. The torque is the difference of the angular momentum current $I^s$ between the far left and far right of the domain wall; the force is the difference of the linear momentum current $T^{11}$ between them. First, for the circularly-polarized incoming wave, $u (\zeta, t) = a \cos (k \zeta - \omega t)$ and $v (\zeta, t) = - a \sin (k \zeta - \omega t)$, the time-averaged torque and force are given by 
\begin{eqnarray}
\tau &=& \mathcal{T}_\kappa a^2 (1 - t_u t_v \cos \Delta \phi_t - r_u r_v \cos \Delta \phi_r ) k  \, , \label{eq:tau} \\
F &=& \mathcal{T}_\kappa a^2 (r_u^2 + r_v^2) k^2 \, , \label{eq:F}
\end{eqnarray}
where $t_u$ and $t_v$ are respectively the transmission amplitudes of the $u$ and $v$ components (and similarly $r_u$ and $r_v$ for the reflection amplitudes) and $\Delta \phi_t$ is the relative phase shift of the transmitted $u$ and $v$ components (and similarly $\Delta \phi_r$ for the reflected wave). Equation~(\ref{eq:tau-cp}) for $\tau$ is the reflectionless limit of Eq.~(\ref{eq:tau}), corresponding to $k \lambda \gg 1$. Secondly, for the linearly-polarized incoming wave, $v (\zeta, t) = a \cos(k \zeta - \omega t)$, the torque vanishes and the force is given by
\begin{equation}
F = \mathcal{T}_\kappa a^2 r_v^2 k^2 \, .
\end{equation}

\emph{Ferromagnetic domain wall.}|The dynamics of the ferromagnet is described by the Lagrangian \cite{Altland2006}, $L = s \int d\zeta \, \mathbf{a}(\mathbf{n}) \cdot \mathbf{n} - U [\mathbf{n}]$, where $\mathbf{a} (\mathbf{n})$ is a vector potential of a magnetic monopole, $\boldsymbol{\nabla}_\mathbf{n} \times \mathbf{a} = \mathbf{n}$. The angular and linear momenta of the domain wall [Eqs.~(\ref{eq:dw})] are given by, respectively, $J = J_0 + 2 s Z$ and $P = P_0 - 2 s \Phi$, where $J_0$ and $P_0$ are arbitrary \cite{YanPRB2013, TchernyshyovAP2015}. Viscous losses can be represented by the Rayleigh dissipation function \cite{Goldstein} $R = \alpha s \int d\zeta \, \dot{\mathbf{n}}^2 / 2$, where $\alpha$ is Gilbert's damping constant \cite{GilbertIEEE2004}. By plugging the domain-wall solution, we obtain $R = \alpha s (\lambda \dot{\Phi}^2 + \dot{X}^2 / \lambda)$. The conservations of the total angular and linear momenta yield 
\begin{equation}
\tau = \dot{J} + 2 \alpha s \lambda \dot{\Phi} \, , \quad F = \dot{P} + 2 \alpha s \dot{Z} / \lambda \, .
\label{eq:tau-F}
\end{equation}
The steady-state velocity $V = \dot{Z}$ is given by
\begin{equation}
V = \frac{\tau + \alpha \lambda F }{ 2 (1 + \alpha^2) s } \, .
\label{eq:V-fm}
\end{equation}

\emph{Antiferromagnetic domain wall.}|The dynamics of the antiferromagnet is described by the Lagrangian, $L = \chi \int d\zeta \, \dot{\mathbf{n}}^2 / 2 - U[\mathbf{n}]$, where $\chi$ quantifies inertia of the order parameter \cite{BaryakhtarFNT5, *AndreevSPU1980}. The Rayleigh dissipation function is $R = \alpha s\int d\zeta \, \dot{\mathbf{n}}^2 / 2$ \cite{GomonayPRB2010}. For slow dynamics, the angular and linear momenta of the domain wall are respectively given by $J = I \dot{\Phi}$ and $P = M \dot{Z}$, where $I \equiv 2 \chi \lambda$ and $M \equiv 2 \chi / \lambda$ are the moment of inertia and the mass of a static domain wall \cite{KimPRB2014}. Their equations of motion are same as Eqs.~(\ref{eq:tau-F}). The steady-state velocity $\dot{Z} (t) \rightarrow V$ is given by
\begin{equation}
V = \frac{\lambda F}{2 \alpha s} \, .
\label{eq:V-afm}
\end{equation}

\emph{Discussion.}|For experiments, linearly polarized elastic waves can be coherently excited by attaching a piezoelectric transducer to the magnetic wire, as done in Refs.~\cite{MatthewsPRL1962, *SytchevaPRB2010} to probe the magnetoacoustic Faraday effect \cite{KittelPR1958}. Coherent excitation of circularly-polarized waves can be generated, for example, by exciting two linearly polarized modes with the fixed relative phase of $\pi / 2$.

Let us make a quantitative estimate for the speed of the domain-wall in ferromagnets, $V = \tau / 2 s$ [Eq.~(\ref{eq:V-fm})] in the zero-damping limit $\alpha = 0$ at the maximum efficiency of the phononic torque, i.e., $\tau = 2 \mathcal{T}_\kappa a^2 k$ [Eq.~(\ref{eq:tau-cp})]. We take the parameters of iron for the magnet, the saturation magnetization $M_s = 2 \times 10^6$ A/m and the mass density $\rho = 7 \times 10^3$ kg/m$^3$ \cite{JilesBook1991}, and the parameter of lead zirconate titanate for the piezoelectric strain constant $d = 10^{-10}$ m/V \cite{BryanPRB2012}. For the iron wire of the cross-sectional area $\mathcal{A} = 20$ nm$^2$ \cite{LipertAPL2010} subjected to the tension $\mathcal{T} = 10^{-3}$ N, the application of the electric field $E = 1$ V/mm rotating at the frequency $10$ MHz across the piezoelectric transducer of length $L = 100$ nm yields the speed $V \approx 40$ m/s (assuming the perfect coupling of stress between the transducer and the wire), which is comparable to the domain-wall speed by the spin-polarized electric current \cite{ParkinScience2008} and by the magnon current \cite{YanPRL2011}.

We have neglected the magnetoacoustic Faraday effect \cite{MatthewsPRL1962} as well as its inverse effect \cite{TokmanEPJB2013}, which are absent in the static treatment of the magnetization according to the energy $U_\text{m}$~(\ref{eq:Um}) that is even under magnetization reversal. The effects might be present when the magnetization is made dynamic, but it is suppressed when the magnetic and the acoustic resonances are significantly mismatched \cite{KittelPR1958}. These effects, in principle, can influence the domain-wall motion. In particular, via the inverse effect, the circularly polarized elastic waves can induce the effective magnetic field along the axial direction and drive the ferromagnetic domain wall by contributing to the force $F$ in Eq.~(\ref{eq:V-fm}) \cite{SchryerJAP1974}. In the zero-damping limit $\alpha = 0$, however, this contribution can be neglected.

\begin{acknowledgments}
We are grateful for discussions with H\'{e}ctor Ochoa and Ricardo Zarzuela, who helped us to formulate the problem. We also thank the anonymous referees, whose comments and questions led to the significant improvement of the Letter. This work was supported by the Army Research Office under Contract No. 911NF-14-1-0016.
\end{acknowledgments}

\bibliographystyle{/Users/evol/Dropbox/School/Research/apsrev4-1-nourl}
\bibliography{/Users/evol/Dropbox/School/Research/master}

\end{document}